\documentstyle[preprint,aps]{revtex}

\begin{document}

\draft

\preprint{YITP-96-19}

\title{Are nontopological strings produced\\
at the electroweak phase transition ?}

\author{Michiyasu Nagasawa
\footnote{Electronic address: nagasawa@yukawa.kyoto-u.ac.jp}}

\author{Jun'ichi Yokoyama}

\address{Yukawa Institute for Theoretical Physics,
Kyoto University, Kyoto 606-01, Japan}

\maketitle

\begin{abstract}
We formulate a local condition for a nontopological defect
to be present. We apply it for electroweak strings and estimate
the probability of their existence at the Ginzburg temperature.
As a result we find strings long enough to serve for baryon-number 
generation are unlikely to be produced.
\end{abstract}


Topological defects are produced at cosmological phase
transitions if vacuum structure after the symmetry breaking
is nontrivial\cite{Kib,topdef}. Even when it is trivial,
however, nontopological defects might be produced.
One of the well-known examples is an electroweak string\cite{ews}.
It has a string-like configuration of the false vacuum which
satisfies field equations of the minimal standard electroweak
model, although whether it constitutes a local energy minimum
is still under investigation\cite{stab}.
While topologically
stable strings have also  been proposed under 
the non-standard extension of the theory
\cite{DS}, we concentrate on the possibility of
nontopological strings within the standard model here.

The electroweak strings might be useful for baryogenesis
in our Universe\cite{BD,naga}.
They can generate an out-of-equilibrium
state even if the electroweak phase transition is of the second
order. Moreover the
electroweak strings themselves have baryon number and may
contribute to the baryon asymmetry production\cite{ewsbn} or
they can induce baryon-number fluctuations through
interaction with background electromagnetic fields\cite{Barr}.
Their effect on the sphaleron transition rate has been
discussed in \cite{Soni}.

All the above analyses, although interesting, rely on
the assumption that the nontopological strings are indeed
produced at the electroweak phase transition more or less
in a similar manner to ordinary topological strings.
However, a more careful analysis is required, since there is no
topological reason for electroweak strings to extend without an end
but they may have a finite length with a monopole-like
configuration at one end and an antimonopole-like configuration
at the other.
Although much work has been done on the stability of 
the width of 
an infinitely long electroweak string\cite{stab}, 
no one has really estimated
their formation rate at the phase transition except for a
preliminary treatment \cite{HDB} in which the authors concerned
mostly with the validity of the geodesic rule in the transient
region between different phases. But their approach is
inappropriate to apply for the present problem, since the
number density of the electroweak strings
cannot be calculated only by the phase distribution of the Higgs
field since a nonvanishing winding number alone does not guarantee
the existence of a false vacuum region and it must be imposed
as an extra condition.
Even if the infinitely long string solution is stable against
perturbation on its width, we cannot say strings are indeed produced
at the electroweak epoch.  Such stability may help
their survival after formation, but their initial number density
must be determined by the realization probability of string-like
configuration at the phase transition. 
In the present {\it Letter} we estimate the formation probability
of the electroweak strings, along which the Higgs fields have
a vanishing amplitude, at the end of the phase transition.

First, for comparison, let us consider the case of an ordinary
topological cosmic string which is produced when local U(1)
symmetry breaks down. In this model, the Higgs field, $\Phi$,
is a complex scalar written by
\begin{equation}
\Phi =\Phi_1 +i\Phi_2\ , \label{eq:phi2}
\end{equation}
where $\Phi_1$ and $\Phi_2$ are real. As is well known, if the
phase of $\Phi$ is randomly distributed on each correlated region,
there should be 0.25 string per one correlation volume\cite{Prok}.
This method, however, cannot be applied to the case of the
nontopological electroweak string since even if any winding number
around a certain region exists, this does not necessarily imply that
a false vacuum is trapped in it. Therefore we start with discussing
the condition for a gauged U(1)
string to be present without resorting to such topological consideration.

The cosmic string can be regarded as a line-like region where
the amplitude of $\Phi$ equals zero. Thus the condition that
a string exists at a certain point in the universe,
$\vec{x}=\vec{a}$, is $\Phi_j\left(\vec{a}\right)=0 ~(j=1,2)$
and at the same time there exists a neighboring point,
$\vec{a}+\vec{\varepsilon}$, where
$\Phi_j\left(\vec{a}+\vec{\varepsilon}\right)=0 ~(j=1,2)$ hold, too.
Since we can always set one of the components of the Higgs field
equal to zero at $\vec{x}=\vec{a}$ using a gauge transformation,
the first condition reduces to having the other component to be
zero, too. On the other hand, $|\vec{\varepsilon}|$ is small by
definition, so the second condition may be rewritten as
\begin{equation}
\Phi_j\left(\vec{a}+\vec{\varepsilon}\right) =\Phi_j\left(\vec{a}\right)
+\vec{\varepsilon}\cdot\vec{\nabla}\Phi_j\left(\vec{a}
\right)=\vec{\varepsilon}\cdot\vec{\nabla}\Phi_j\left(\vec{a}\right)=0
~(j=1,2)\ ,
\label{eq:expan}
\end{equation}
that is, there should exist a spatial vector $\vec{\varepsilon}$
orthogonal to both $\vec{\nabla}\Phi_1\left(\vec{a}\right)$ and
$\vec{\nabla}\Phi_2\left(\vec{a}\right)$. But one can always find such
a vector simply by choosing a normal vector to the plane defined by
$\vec{\nabla}\Phi_1\left(\vec{a}\right)$ and $\vec{\nabla}\Phi_2
\left(\vec{a}\right)$.
Thus once we find $\Phi_j=0$ at $\vec{x}=\vec{a}$, a line-like
configuration of the false vacuum extends without an end,
which is a consequence of the topological structure of the vacuum
manifold of the Abelian Higgs model.

Before proceeding to the case of the electroweak string, here
we consider a rather inconceivable possibility of domain wall
formation in the above model. A domain wall configuration can be
easily shown to exist as a nontopological defect in this model,
for example, by a distribution such that $\Phi_2\left(\vec{x}\right)=0$
everywhere and that $\Phi_1\left(\vec{x}\right)$ obeys a similar
solution as a domain wall in a model of a real scalar field.

The condition that a domain wall exists at $\vec{x}=\vec{a}$ is,
in addition to having $\Phi_j\left(\vec{a}\right)=0$, there should
exist two linearly independent spatial vectors, $\vec{\varepsilon}_1$
and $\vec{\varepsilon}_2$, which satisfy
$\Phi_j\left(\vec{a}+\vec{\varepsilon}_n\right)=0 ~(n=1,2)$ or
\begin{equation}
\vec{\varepsilon}_n\cdot\vec{\nabla}\Phi_j\left(\vec{a}\right)=0~
(n=1,2;~j=1,2)\ .   \label{eq:condw}
\end{equation} 
The necessary and sufficient condition for it is that
$\vec{\nabla}\Phi_1\left(\vec{a}\right)$ and
$\vec{\nabla}\Phi_2\left(\vec{a}\right)$ are parallel to each other
including the trivial case that one or the both of them have a
vanishing amplitude.
We can also show that the above condition is gauge-invariant.
In fact, since gauge-transformation is a linear transformation
for the Higgs fields such as
\begin{equation}
\Phi'_j\left(x\right)=\sum_l c_{jl}\left(x\right)\Phi_l\left(x\right)\ ,
\end{equation}
we find
\begin{equation}
\vec{\nabla}\Phi'_j\left(x\right)=
\sum_l \vec{\nabla} c_{jl}\left(x\right)\cdot\Phi_l\left(x\right)
+\sum_l c_{jl}\left(x\right)\vec{\nabla}\Phi_l\left(x\right)\ ,
\end{equation}
but at $\vec{x}=\vec{a}$ we have $\Phi_l\left(\vec{a}\right)=0$
by assumption, so
\begin{equation}
\vec{\nabla}\Phi'_j\left(\vec{a}\right)=
\sum_l c_{jl}\left(\vec{a}\right)\vec{\nabla}\Phi_l\left(\vec{a}\right)\ ,
\end{equation}
which implies the gauge-invariance of (\ref{eq:condw}).
Furthermore we can choose the spatial coordinate at $\vec{x}=\vec{a}$
such that $\vec{\nabla}\Phi_1\left(\vec{a}\right)$ has a nonvanishing
component only in the $x$-component. Then the condition
(\ref{eq:condw}) reduces to
\begin{equation}
\partial_y \Phi_2\left(\vec{a}\right)=
\partial_z \Phi_2\left(\vec{a}\right)=0\ .
\end{equation}
Thus in this case two additional conditions must be satisfied to
produce a nontopological defect.

Now we return to the electroweak string. In the minimal standard model,
the Higgs field, $\phi$, is an SU(2) doublet and we write it as
\begin{equation}
\phi = \left(
\begin{array}{c}
\phi_1+i\phi_2 \\ \phi_3+i\phi_4
\end{array}
\right)\ ,
\end{equation}
where $\phi_j ~(j=1,2,3,4)$ is a real component. Similarly to the
case of the Abelian Higgs model, the conditions for the existence
of a string at $\vec{x}=\vec{a}$ are that
$\phi_j\left(\vec{a}\right)=0 ~(j=1,2,3,4)$ and that there exist
an infinitesimal spatial vector $\vec{\varepsilon}$ such that
$\phi_j\left(\vec{a}+\vec{\varepsilon}\right)=0 ~(j=1,2,3,4)$.
Since we can rotate $\phi$ using a gauge transformation
so that only one component is nonvanishing at $\vec{x}=\vec{a}$,
the first condition reduces to that the remaining component is
also vanishing, whose probability is denoted by $p_0$ hereafter.
The second condition reads
\begin{equation}
\vec{\varepsilon}\cdot\vec{\nabla}\phi_j\left(\vec{a}\right)=0~(j=1,2,3,4)\ ,
\label{eq:cond4}
\end{equation}
which can again be shown to be gauge-invariant.

For a nontrivial solution of $\vec{\varepsilon}$ to exist, it is
necessary and sufficient that all the vectors
$\vec{\nabla}\phi_j\left(\vec{a}\right)$ lie in the same plain
defined by two linearly independent vectors, say,
$\vec{\nabla}\phi_1\left(\vec{a}\right)$ and
$\vec{\nabla}\phi_2\left(\vec{a}\right)$. Then a normal vector to
that plain can serve as $\vec{\varepsilon}$ and the remaining
conditions turns out to be
\begin{equation}
\vec{\varepsilon}\cdot\vec{\nabla}\phi_3\left(\vec{a}\right)=0\quad {\rm and}\quad
\vec{\varepsilon}\cdot\vec{\nabla}\phi_4\left(\vec{a}\right)=0\ .
\label{eq:orths}
\end{equation}
Now we can set the spatial coordinate so that the normal vector
to the plain, $\vec{\varepsilon}$, has a nonvanishing component
only along the $x$-direction. Then the conditions
(\ref{eq:orths}) reduce to
\begin{equation}
\partial_x \phi_3\left(\vec{a}\right)=0\quad {\rm and}\quad
\partial_x \phi_4\left(\vec{a}\right)=0\ .
\end{equation}
Assuming that $\partial_x \phi_3$, $\partial_x \phi_4$ and
the amplitude of the Higgs field behave independently and denoting
the probability of having $\partial_x \phi_j\left(\vec{a}\right)=0$
by $d_0$, the probability, $P_s$, that there exist a string-like
false vacuum region in the infinitesimal neighborhood at
$\vec{x}=\vec{a}$ turns out to be
\begin{equation}
P_s \sim p_0 d_0^2\ .
\end{equation}
This is smaller than the case of ordinary topological strings
at least by the factor of $d_0^2$.
Obviously we can predict that the more components
the Higgs field has, the more difficult it becomes to produce
a string, with the higher power of $d_0$.

For the purpose of estimating $p_0$ and $d_0$, we introduce
the probability distribution function (PDF) of the Higgs field
in the thermal bath.
We employ the Hartree approximation\cite{BVHLS}
with which the higher moment of the field can be described by
the second moment. Then the amplitude of a scalar field, $\phi$,
obeys a random Gaussian probability distribution such as
\begin{equation}
P_\phi\left( \phi \right)d\phi =\frac{1}{\sqrt{2\pi}\sigma}
\exp\left\{-\frac{\left(\phi-c\right)^2}{2\sigma^2}\right\}d\phi\ ,
\label{eq:pphi}
\end{equation}
where $c$ is the the averaged value of $\phi$ and $\sigma$ is
the standard deviation. Under the same assumption, the gradient
of the Higgs field component obeys the PDF
\begin{equation}
P_{\partial\phi}\left( \partial_l\phi_j \right)d\left(\partial_l\phi_j
\right) =\frac{1}
{\sqrt{2\pi}\eta}\exp\left\{-\frac{\left(\partial_l\phi_j\right)^2}
{2\eta^2}\right\}d\left(\partial_l\phi_j\right)\ ,
\end{equation}
where the averaged value of $\partial_l\phi_j$ equals zero.
The dispersion, $\eta$ , which is independent of $l$, can be written as
\begin{equation}
\eta^2 =\frac{1}{6\pi^2}\int P\left(k\right)k^4dk\ ,
\label{eq:eta}
\end{equation}
together with
\begin{equation}
\sigma^2 =\frac{1}{2\pi^2}\int P\left(k\right)k^2dk\ ,
\label{eq:sigma}
\end{equation}
where $P(k)$ is the power spectrum, or the Fourier transform of 
$\left<\phi(\vec{0})\phi(\vec{x})\right>-c^2$\cite{BBKS}. 
Using these formulae, $p_0$ and $d_0$ can be written as
\begin{equation}
p_0=P_\phi\left(0\right)\cdot\delta\phi\ ,\quad
d_0=P_{\partial\phi}\left(0\right)\cdot\delta\partial\phi\ , \label{kakuritu}
\end{equation}
where $\delta\phi$ and $\delta\partial\phi$ are some width scales.
The values of $\sigma$ and $\eta$ can be obtained by substituting
\begin{equation}
P(k)=\left(k^2+m_0^2\right)^{-1/2}\left(e^{\frac{\sqrt{k^2+m_0^2}
}{T}}-1\right)^{-1}
\label{eq:pk}
\end{equation}
to the equations (\ref{eq:sigma}) and (\ref{eq:eta})
where $m_0^2$ is the effective mass squared
at $\phi=c$\cite{DLHLL}.

In the standard
electroweak theory, the one-loop 
effective potential for the Higgs field
with the finite temperature corrections is written as\cite{DLHLL,ewpt}
\begin{equation}
V_{\rm eff}\left( \phi \right) = D\left( T^2-T_2^2\right)\phi^2
-ET\phi^3+\frac{\lambda_T}{4}\phi^4\ , \label{eq:potew}
\end{equation}
where $T_2$ is the temperature when the symmetric state, $\phi=0$,
becomes unstable. Using the standard values of the parameters such
as $m_W=80.6$ GeV for the W-boson mass, $~m_Z=91.2$ GeV for
the Z-boson mass, and $~m_t=174$ GeV for the top quark mass,
the coefficients in the potential (\ref{eq:potew}) are calculated as
$D=0.169$, $E=0.00965$, $T_2=92.6, 134.3, 249.8$GeV, and 
$\lambda_{T=T_2}=0.0354, 0.0747, 0.300$,
when the mass of the Higgs particle is
$m_H=60, 100, 200$ GeV, respectively. 

We estimate the string formation at the Ginzburg temperature,
$T=T_G$, when the defects are considered to turn stable against
thermal fluctuations\cite{Kib,tdf}. $T_G$ is evaluated by the
condition, $T=\Delta V\xi^3$, where $\Delta V$ is the potential-energy
density gap between the symmetric state and the potential minimum and
$\xi$ is the correlation scale of $\phi$, which is defined by the
square-root inverse of the second derivative of the effective
potential at its minimum. Numerically we find
\begin{equation}
T_G=76.9,~62.4,~34.4~~{\rm GeV}\ ,
\end{equation}
for $m_H=60, 100, 200$ GeV, respectively. Thus $T_G$ is always
smaller than $T_2$, which implies that even if the electroweak
phase transition might start as a first-order transition its
final stage is described by the dynamics of a second-order phase
transition as far as defects formation is concerned.

In the Hartree approximation, 
the potential (\ref{eq:potew}) is simplified using the replacement
\begin{equation}
\varphi^3 \longrightarrow 3\sigma^2\varphi\ ,\quad
\varphi^4 \longrightarrow 6\sigma^2\varphi^2-3\sigma^4\ ,
\label{eq:hart}
\end{equation} 
where $\varphi \equiv \phi-c$ and $\sigma$ is the root mean square
of $\varphi$ which should be equal to the standard deviation in
the equation (\ref{eq:pphi}).
At $T=T_G$, we obtain the effective mass of $\varphi$ from the
coefficient of the quadratic 
term in the approximate potential as
\begin{eqnarray}
m^2&\equiv& m_\varphi^2+\delta m^2\ , \label{eq:emass}\\
m_\varphi^2&=&2D\left(T_G^2-T_2^2\right)-6ET_Gc+3\lambda_{T=T_G}c^2~~,~~
\delta m^2=3\lambda_{T=T_G}\sigma^2\ .
\end{eqnarray}
In order that 
the expectation value of $\varphi$ vanishes, or 
$\varphi$ has its potential minimum at $\varphi=0$,
the consistency condition for $c$,
\begin{equation}
2D\left(T_G^2-T_2^2\right)c-3ET_G\left(c^2+\sigma^2\right)
+\lambda_{T=T_G}c\left( c^2+3\sigma^2\right)=0\ , \label{eq:center}
\end{equation}
must be satisfied. 
Now we substitute $m$ into $m_0$ in (\ref{eq:pk})
and then numerically solve equations (\ref{eq:sigma}), (\ref{eq:emass})
and (\ref{eq:center}) in a self-consistent manner.
Using $\lambda_{T=T_G}=0.0422, 0.103, 0.372$, we find
\begin{eqnarray}
c &=& 172.4,~224.3,~237.2~~{\rm GeV}\ ,\\
\sigma &=& 17.1,~9.05,~0.717~~{\rm GeV}\ ,\\
m &=& 46.1,~99.6,~204~~{\rm GeV}\ , \label{eq:mass}
\end{eqnarray}
and equation (\ref{eq:eta}) yields
\begin{equation}
\eta =1860,~995,~67.5~~{\rm GeV}^2\ .
\end{equation}
Here and hereafter, all the numerical values correspond to the cases
$m_H=60, 100, 200$ GeV, respectively. We can see that
$\delta m^2 \ll m_\varphi^2$
justifies the Hartree approximation (\ref{eq:hart}).

$p_0$ and $d_0$ are explicitly calculated as
\begin{eqnarray}
p_0&=& \left(2.6\times 10^{-23},~1.8\times 10^{-134},~
10^{-54682}\right)\times \alpha \ ,\\
d_0&=& \left(0.168,~0.362,~0.865\right)\times \frac{\alpha}{\beta}\ ,
\end{eqnarray}
where we have put $\delta\phi =\alpha \sigma$ and $\delta\partial\phi
=\alpha\sigma/\beta m^{-1}$
with $\alpha$ and $\beta$ being constant.
That is, we have normalized $\delta\phi$ by their  variance
and $\delta\partial\phi$ by $\delta\phi$ divided by the correlation
length $\xi=m^{-1}$.
Thus the probability to find a string stretched from  $\vec{x}=\vec{a}$ to
$\vec{x}=\vec{a}+\vec{\varepsilon}$ is as small as
\begin{equation}
P_s \sim p_0 d_0^2 \sim \left( 7.5\times 10^{-25},~2.3\times 10^{-135},~
10^{-54682}\right)\times \frac{\alpha^3}{\beta^2} \ .
\end{equation}
Since $\sigma \ll c$ holds already at the Ginzburg temperature,
$p_0$, which is calculated as (\ref{kakuritu}), turns out to be
extremely small. This, however, might not be a fatal
problem itself.  If false vacuum defects decouple from thermal
equilibrium at a higher temperature, say, when $\sigma$ becomes
smaller than $c$, we should estimate the probability at that
temperature.  Then $P_\phi (0)$, which is very sensitive to the
temperature, could be  larger. 
More serious
is the extra suppression factor for a string to extend for
a finite length $\epsilon=|\vec{\varepsilon}|$, $d_0^2$, which is 
less sensitive to the temperature.  For example,
for a string to extend for the correlation length 
\begin{equation}
\xi=m^{-1}= 0.022,~0.010,~0.0049~{\rm GeV}^{-1}\ ,
\end{equation}
$d_0^2$ is as small as 
\begin{equation}
d_0^2=\left(0.028,~0.13,~0.75
\right)\times\left(\frac{\alpha}{\beta}\right)^2\ ,
\end{equation} 
respectively. But this is not the whole story. Since
the discussion based on the lowest-order expansion of
$\phi\left(\vec{a}+\vec{\varepsilon}\right)$ 
is valid only if the inequality
\begin{equation}
\max\left( \delta\phi, \epsilon\delta(\partial\phi) \right)=
\max\left( \alpha\sigma, \epsilon\frac{\alpha\sigma}{\beta
m^{-1}}\right)
> |\varepsilon_i\varepsilon_j\partial_i\partial_j\phi|
\sim \epsilon^2 \sqrt{\left<\left(\partial^2\phi\right)^2\right>}\ ,
\end{equation}
is satisfied. Calculating the root-mean square of $\partial^2\phi$
in the same way as in (\ref{eq:eta}), we find
\begin{equation}
\epsilon \lesssim \max\left( (0.0059,~0.0062,~0.0078)\sqrt{\alpha},~~
(0.0016,~0.0039,~0.012)\frac{\alpha}{\beta} \right)
~~{\rm GeV}^{-1}\ ,
\end{equation}
which is smaller than or barely comparable to the correlation
length of the Higgs field for 
$\alpha,~\beta \lesssim {\cal O}(1)$. 
Therefore for a string to extend for the correlation length,
we must impose constraints on the amplitude of higher derivatives
of $\phi$ as well, which results in further suppression factor
in the formation probability.

One may wonder that in the case of  stable nontopological strings 
there may be a correlation between $\phi=0$ and $\partial\phi=0$
and that we may have a larger probability of their formation.
However, previous stability analyses of electroweak strings are all
concerned with that of the string width of an infinitely long string
solution\cite{stab}, 
while strings with finite length are unstable and tend to
shrink\cite{BD}.
Thus the solidness of the string core alone does not help to
realize a long string-like configuration. 
We therefore conclude it is very difficult
to find a string longer than the correlation length.
In other words, even if a false vacuum string is produced,
its length is comparable to its width, and such a configuration
should be called a false vacuum ball rather than a string.
Thus we cannot make use of such objects for baryogenesis.

In summary, we have considered how difficult it is to produce
nontopological defects by the Kibble mechanism in cosmological
phase transitions. As a specific example, we have discussed that
electroweak strings which are long enough to serve for baryogenesis
are very unlikely to be present at the Ginzburg temperature when
the defects become stable against thermal fluctuations.

This work was partially supported by the Japanese Grant
in Aid for Science Research Fund of the Ministry of Education, Science
and Culture (Nos. 5110, 08740202).

\end{document}